\documentclass[11pt]{article}
\usepackage{xspace}
\usepackage{nsd}
\newcommand{\occ}{\mathrm{occ}}
\let\Im\relax
\DeclareMathOperator{\Im}{Im}
\newcommand{\E}{\expect}

\newcommand{\FI}{\problem{Function Inversion}}
\newcommand{\FIRep}{\problem{Function Inversion Reporting}}
\newcommand{\FICount}{\problem{Function Inversion Counting}}

\newcommand{\ThreeSUMInd}{\problem{3SUM-Indexing}}
\newcommand{\ThreeSUMIndRep}{\problem{3SUM-Indexing Reporting}}

\newcommand{\kSUMInd}{\problem{$k$SUM-Indexing}}
\newcommand{\kSUMIndRep}{\problem{$k$SUM-Indexing Reporting}}
\newcommand{\ThreeXORIndRep}{\problem{3XOR-Indexing Reporting}}

\newcommand{\kXORIndRep}{\problem{$k$XOR-Indexing Reporting}}

\newcommand{\GSIRep}{\problem{Gapped String Indexing Reporting}}

\newcommand{\JIRep}{\problem{Jumbled Indexing Reporting}}
\newcommand{\RangeSearching}{\problem{Range Searching}}
\newcommand{\RangeCounting}{\problem{Range Counting}}
\newcommand{\RangeRep}{\problem{Range Reporting}}
\newcommand{\ANN}{\problem{Approximate Nearest Neighbor}}

\theoremstyle{plain}
\newtheorem{maintheorem}{Theorem}
\newtheorem{mainproposition}[maintheorem]{Proposition}

\date{}
\title{Function Inversion Reporting}
 \author{
     Itai Dinur
     \thanks{Ben-Gurion University and Georgetown University. Email: \texttt{dinuri@bgu.ac.il}. Supported by a gift to Georgetown University.}
     \and
 	Alexander Golovnev
 	\thanks{Cornell University. Email: \texttt{alexgolovnev@gmail.com}. Supported by the National Science Foundation CAREER award (grant CCF-2338730).}
     \and
     Chao Yan
     \thanks{Technion. Email: \texttt{cy399@georgetown.edu}. Funded by the European Union (ERC, GENERALIZATION, 101039692). Views and opinions expressed are however those of the author(s) only and do not necessarily reflect those of the European Union or the European Research Council Executive Agency. Neither the European Union nor the granting authority can be held responsible for them.}
     \and
     Yiding Zhu
     \thanks{Georgetown University. Email: \texttt{yz1464@georgetown.edu}.}
 }

\begin{document}
\maketitle   
\begin{abstract}
In the \FI problem, the goal is to invert a function $f:[N] \to [N]$ in two phases. In the preprocessing phase, the offline algorithm computes an advice string of length $S$ bits. In~the online phase, the online algorithm receives the advice and a challenge $y \in [N]$, and the goal is to find a preimage $x \in [N]$ such that $f(x) = y$ in time $T$. The problem was initially studied in cryptography, but in recent years it has found various applications to data structure problems. 
Yet, several of the applications require a variant of \FI, known as \FIRep, where the goal is to output all preimages of the challenge~$y$. 

In this paper, we study the \FIRep problem in a systematic way by unifying and improving existing  algorithms. As a result, we obtain improved data structures for several problems including \ThreeSUMIndRep, \RangeRep, \ANN, \JIRep, and \GSIRep.
\end{abstract}
\mytableofcontents

\section{Introduction}
\FI is a fundamental problem in cryptography, where the goal is to invert an efficiently computable function $f:[N] \to [N]$ in two phases. 
In the preprocessing phase, the offline algorithm~$\A_0$ is given access to $f$ and computes an advice string of length $S$ bits. In the online phase, the online algorithm $\A_1$ receives the advice string and a challenge $y \in [N]$ and the goal is to find a preimage $x \in [N]$ such that $f(x) = y$ in time $T$ (assuming such a preimage exists). The problem was first studied by Hellman~\cite{H80} who gave a heuristic algorithm for the problem with a time-space tradeoff of $T S^2 = \widetilde{O}(N^2)$,\footnote{The $\widetilde{O}$ notation suppresses factors poly-logarithmic in the input length.} assuming that $f$ is chosen uniformly at random.

Fiat and Naor~\cite{FN91} made Hellman's algorithm rigorous by considering an arbitrary function $f: [N] \to [N]$ with collision probability $q(f) := \Pr_{x,x' \sim [N]} [f(x) = f(x')]$, and extended the time-space tradeoff for such a function to $T S^2 = \widetilde{O}(N^3) \cdot q(f)$. Moreover, in the worst case, the algorithm of~\cite{FN91} has a time-space tradeoff of $T S^3 = \widetilde{O}(N^3)$.

Despite originating in cryptography,
in the past few years, the \FI problem has found many applications in state-of-the-art data structures~\cite{CK19,KP19,GGHPV20,AESZ23,BGLPRS24,ACDI24,M24,DG26,BBM26,KKMV26,CEEGS26}. A closer inspection shows that several applications~\cite{ACDI24,BGLPRS24,M24,BBM26,KKMV26} use \FIRep, which is a variant of \FI where the goal of the online algorithm is to find \emph{all} preimages $x$ such that $f(x) = y$.
Given the increasing number of applications, the goal of this paper is to address the \FIRep problem in a unified way.

A simple algorithm for \FIRep uses the Fiat-Naor algorithm in a black-box way to solve it with the tradeoff $S^3 T = \widetilde{O}(N^3) \cdot (\occ(y)+1)$, where $\occ(y) := |f^{-1}(y)| = |\{x : f(x) =y\} |$. This algorithm was applied (implicitly) in~\cite{ACDI24,BGLPRS24,M24}.
\begin{mainproposition}[Folklore]\label{prop:GeneralReporting}
For every $1\leq S\leq N$, there exists a data structure that preprocesses an arbitrary function $f \colon [N] \to [N]$ into space~$\widetilde{O}(S)$, and for any query $y \in [N]$, it outputs all $\occ(y)$ preimages of $y$ in time~$\widetilde{O}(T)$ for 
    \[
     T = \frac{N^3}{S^3} \cdot (\occ(y)+1)\;.
    \]
\end{mainproposition}
We now sketch the proof. Let $2^{i-1} \leq \occ(y) < 2^i$ and $K=2^i \cdot \poly\log(N)$. Randomly partition the domain into $K$ parts, each of size $N / K$. This gives us functions $f_1,\ldots,f_K$ with domain size $N'=N/K$. With high probability, each domain contains (at most) one preimage of $y$, hence it is sufficient to solve \FI for all $K$ functions and report all found preimages. Applying the Fiat-Naor algorithm to each function,\footnote{The Fiat-Naor algorithm can be easily adapted to functions with different domain and range sizes. The tradeoff remains $T S^3 = \widetilde{O}(N^3)$, where $N$ is the domain size.} costs time $T'$ and space $S'$ satisfying $S'^3 T'= \widetilde{O}(N'^3)$. Since $S=S' K$ and $T=T' K$, we can solve the original reporting problem with the tradeoff $S^3 T=\widetilde{O}(N^3K)=\widetilde{O}(N^3) \cdot \occ(y)$.

Curiously, in the special case that each $y$ satisfies $|f^{-1}(y)|\leq\Delta$, one can do better. The theorem below was proven in~\cite[Theorem~5]{BBM26} and \cite[Corollary~2.7]{KKMV26}.
\begin{restatable}[{\cite{BBM26,KKMV26}}]{mainproposition}{PropDelta}\label{prop:BoundedReporting}
     Let $f \colon [N] \to [N]$ be a function such that for every $y \in [N]$, $|f^{-1}(y)|\leq\Delta$. For all $1\leq S\leq N$, there exists a data structure that preprocesses $f$ into space $\widetilde{O}(S)$, and for any query $y \in [N]$, it outputs all $\occ(y)$ preimages of $y$ in time $\widetilde{O}(T)$ for
     \[
     T = \frac{N^2}{S^2}\cdot\Delta \;.
     \]
With probability $1-1/N$ over the randomness of the preprocessing stage, the data structure answers all queries correctly, and the preprocessing time is $\widetilde{O}(N)$.
\end{restatable}
Note that the tradeoff of \cref{prop:BoundedReporting} matches the best known tradeoff for $\FI$ in this regime! That is, we get all $\Delta$ preimages for free.

The idea of the data structure in \cref{prop:BoundedReporting} is to use the function-sensitive Fiat-Naor tradeoff of $T S^2 = \widetilde{O}(N^3) \cdot q(f)$. By randomly partitioning the domain into $K=\Delta\cdot \poly\log(N)$ parts, with high probability, we obtain $K$ functions where each $y$ has a unique preimage. In particular, the functions have domains of size $N'=N/K$ and collision probability $q=\widetilde{O}(1/N')$. From the Fiat-Naor algorithm for \FI, we then have the tradeoff of $(S')^2 T' = \widetilde{O}(N'^2)$ for each function, where $S=S' K$ and $T=T' K$.
This gives us the tradeoff of $S^2 T= \widetilde{O}(N^2)\cdot \Delta$ for $\FIRep$. 

\subsection{Our Results}

Although~\cref{prop:BoundedReporting} does not give a general-case \FIRep algorithm, it suggests that the tradeoff of~\cref{prop:GeneralReporting} may not be optimal in general. Indeed, in this work, we show that~\cref{prop:GeneralReporting} can be improved in nearly all regimes.
\begin{restatable}{maintheorem}{ThmMain}\label{thm:main}
 For every $1\leq S\leq N$, there exists a data structure that preprocesses an arbitrary function $f \colon [N] \to [N]$ into space~$\widetilde{O}(S)$, and for any query $y \in [N]$, it outputs all $\occ(y)$ preimages of $y$ in time~$\widetilde{O}(T)$ for 
    \[
    T = \frac{N^3}{S^3} + \frac{N^2}{S^2}\cdot \occ(y)\;.
    \]
With probability $1-1/N$ over the randomness of the preprocessing stage, the data structure answers all queries correctly, and the preprocessing time is $\widetilde{O}(N)$.
\end{restatable}
Thus,~\cref{thm:main} is better than~\cref{prop:GeneralReporting} if $\occ(y) \geq N^{\epsilon}$ for some $\epsilon > 0$ (except for some trivial edge cases such as $S = \widetilde{\Omega}(N)$).

\cref{thm:main} applies to all functions $f$. Yet, similar to Fiat and Naor~\cite{FN91}, we can obtain a tradeoff that is sensitive to the collision probability $q(f)$.
\begin{restatable}{maintheorem}{ThmGeneral}\label{thm:basic FIR result}
    For every $1\leq S \leq N$, there exists a data structure that preprocesses an arbitrary function $f \colon [N] \to [N]$ into space~$S$, and for any query $y \in [N]$, it outputs all $\occ(y)$ preimages of $y$ in time~$\widetilde{O}(T)$ for 
    \[
    T = \frac{N^3}{S^2} \cdot q(f) + \frac{N}{S} \cdot \occ(y)\;.
    \]
    With probability $1-1/N$ over the randomness of the preprocessing stage, the data structure answers all queries correctly, and the preprocessing time is $\widetilde{O}(N)$.
\end{restatable}
We note that this theorem immediately implies~\cref{prop:BoundedReporting} since $q(f) \leq \frac{\Delta}{N}$. It can also be used to obtain \cref{thm:main}, but gives a better tradeoff than~\cref{thm:main} in various cases (e.g., $q(f) = \tilde{O}\left(1/N \right)$ and $\occ(y) \geq N^{\epsilon}$ for some $\epsilon > 0$). 

Finally, almost all results in this work can be proven directly from \cref{thm:main}. Yet, the function-sensitive~\cref{thm:basic FIR result} gives a more complete treatment of the \FIRep problem and may be used in future applications.

We now briefly demonstrate one application of \cref{thm:main}. In the \ThreeSUMIndRep problem, the goal is to preprocess $n$ integers $a_1,\ldots,a_n$ using space~$S$ so that each query can be answered in time~$T$. A query consists of an integer~$y$, and the answer is the list of all pairs $(i,j)$ such that $a_i+a_j=y$. It was shown in~\cite{KP19,GGHPV20} that it suffices to simply output $f^{-1}(y)$ for the function $f\colon [n]\times [n]\to\Z$ defined by
\[
f(i,j) = a_i+a_j\;.
\]
The function~$f$ has the domain of size $N=n^2$, and, therefore, admits a solution with $T= \frac{n^6}{S^3}\cdot\occ(y)$ using \cref{prop:GeneralReporting}. Prior to this work, this was the best known data structure for \ThreeSUMIndRep~\cite{BGLPRS24}. A straightforward application of \cref{thm:main} to the same function~$f$, immediately improves the query time to $T= \frac{n^6}{S^3}+\frac{n^4}{S^2}\cdot\occ(y)$. In \cref{sec:3sum}, we will further improve this tradeoff in certain parameter regimes.

Similarly, an application of \cref{thm:main} (or of its simple extensions from \cref{sec:extensions}) immediately implies improved bounds for a host of problems from different areas that have previously used \FI. We demonstrate such applications in \cref{sec:apps} for \kSUMIndRep, \JIRep, \GSIRep, \RangeRep, \RangeCounting, and \ANN.

\subsection{Our Techniques}

We informally summarize our main techniques.

\subsubsection{\cref{thm:main}.}
The proof of~\cref{thm:main} is simple. Given $f:[N] \to [N]$, we will carefully derive a logarithmic number of functions from it. Each derived function agrees with $f$ only for inputs $x$ such that $\occ(f(x))$ is lower than some threshold (which increases exponentially in the index of the derived function). For inputs $x$ with $\occ(f(x))$ higher than the threshold, the derived function disperses their images so that their $\occ$ becomes at most 2. We show that this dispersion allows us to apply~\cref{prop:BoundedReporting} to each derived function. In the online phase, given $y$ (with $\occ(y)$ sufficiently large), we first use the precomputed data to compute its $\occ$ exactly, and then search the data structure of the derived function (guaranteed by~\cref{prop:BoundedReporting}) that allows us to report all preimages in minimal time. More specifically, we search the first data structure in the sequence for which the preimages of $y$ are not dispersed. We remark that if $\occ(y)$ is low, its exact value is not directly stored, as there can be many such images $y$. In this case, all the preimages of $y$ are retrieved from the first data structure in the sequence. 

\subsubsection{\cref{thm:basic FIR result}.}
Unlike~\cref{thm:main}, it is not clear how to derive~\cref{thm:basic FIR result} from previous results in a black-box manner. Therefore, we derive it using the Fiat-Naor data structure in a white-box way. In fact, our data structure is somewhat similar to rainbow tables~\cite{Oechslin03}.

For a parameter $\ell \approx N \cdot q(f)$, we will build $\ell$ randomized and independent tables, each with the property that every $x \in [N]$ is covered in a table with probability of about $\frac{1}{\ell}$  (ignoring logarithmic factors). 
Thus, (about) $\ell$ tables suffice to cover all $x \in [N]$ with high probability.

We now focus on the structure of a single table $i \in [\ell]$. Table $i$ contains $m$ chains of length $t$, where $m \approx \frac{S}{N q(f)}$ and $t \approx N/S$ (and thus $\ell m t \approx N$).
The elements of chain $j \in [m]$ are denoted by 
\[X^{i,j}_{0}, X^{i,j}_{1}, \ldots, X^{i,j}_{t}\;,\]
where $X^{i,j}_{0}$ is selected uniformly at random and 
for $0 \leq a \leq t-1$, 
$X^{i,j}_{a+1} = g_{i,a}(  f(X^{i,j}_{a} ))$.
Here, $\{g_{i,a}\}_{i \in [\ell], a \in \{0,\ldots,t-1\}}$ are selected independently and uniformly from a family of pairwise independent hash functions.\footnote{The fact that we need only pairwise independent hash functions (rather than $t$-wise independent hash functions) makes our data structure and proof simpler than those of~\cite{FN91}.} For each chain, we store only its start and end points.

The total space occupied by all the chains in all the tables is about $\ell m \approx N q(f) \frac{S}{N q(f)} = S$, and the total space occupied by all the seeds of the pairwise independent functions is about $\tilde{O}(1) \cdot \ell t \approx q(f)N \frac{N}{S} = \frac{q(f)N^2}{S}$. We may assume that $\frac{q(f)N^2}{S} \lesssim S$. Indeed, if $q(f) \gtrsim \frac{S^2}{N^2}$, then the desired tradeoff gives $T \gtrsim N$, and we can solve the problem trivially in time $\widetilde{O}(T)$. Overall, the space complexity is (about) $S$, as required.

We sketch an argument that shows that each $x \in [N]$ is covered in a table with probability of about $\frac{1}{\ell} \approx \frac{1}{N\cdot q(f)} \approx \frac{mt}{N}$. To show this, it is essentially sufficient to show that there are not too many repetitions among the $mt$ points covered in the table (with multiplicities counted). Specifically, we will show that there are $\Omega(mt)$ distinct points covered in a table on average. For table $i$ we say that two chains $j_1 \neq j_2$ merge if there exists $1 \leq a \leq t$ such that 
$X^{i,j_1}_{a} = X^{i,j_2}_{a}$. Observe that for fixed $a \geq 2$, the probability that $X^{i,j_1}_{a} = X^{i,j_2}_{a}$ assuming
$X^{i,j_1}_{a-1} \neq X^{i,j_2}_{a-1}$ is about $q(f)$ by the pairwise independence of $g_{i,a-2}$. Therefore, the probability that two chains merge is roughly bounded by $t \cdot q(f)$. Thus, a particular chain merges with some other chain with probability roughly $tm \cdot q(f)$. Since $tm \approx 1/q(f)$, we can set the parameters such that, in expectation, at most a constant fraction (which is less than 1) of the chains will merge with some other chain. After showing that other types of collisions (not aligned in a specific $a$) do not contribute much more, this is sufficient to lower bound the number of distinct covered points in a table by $\Omega(mt)$.   

Finally, we consider the online phase. In this phase, given $y$, we search each of the $\ell$ tables for its preimages and output them. To search table $i \in [\ell]$, for each potential position in a chain $0\leq a\leq t$, we compute the continuation of the chain starting from $y$ and check whether the endpoint is stored in table $i$. If so, for each matching endpoint we restart the chain computation from the corresponding start point, hoping to obtain a preimage of $y$. 
The total time complexity can be bounded as follows: first, we bound the time of all chain computations that do not yield any endpoint match.
As each chain computation per $i,a$ takes about $t$ time, the total time is about $\ell t^2 \approx q(f)N \frac{N^2}{S^2} = \frac{N^3}{S^2} \cdot q(f)$. This accounts for the first term in the tradeoff. It remains to account for the second term that bounds the time it takes for all chain computations that yield endpoint matches. We show that the total number of such matches is about $\occ(y)$ (clearly it is at least $\occ(y)$, and we essentially show it is not much more on average). This gives a total time of about $t \cdot \occ(y) \approx \frac{N}{S} \occ(y)$ and accounts for the second term in the tradeoff.

\section{Preliminaries}
We denote the sets of integers and non-negative integers by $\Z$ and $\Z_{\geq0}$, respectively.
For a positive integer $m$, $[m]$ denotes the set of integers $\{1,\ldots,m\}$. 
For two vectors $x,y\in\{0,1\}^\ell$, by $x\oplus y\in\{0,1\}^\ell$ we denote their bitwise XOR. 

We use $\log(\cdot)$ to denote the logarithm base~2, i.e., $\log(2^n) = n$, and we use $\ln(\cdot)$ to denote the natural logarithm.

For a function $f\colon[N]\to[N]$, $\Im(f)$ denotes the image of~$f$. For $y\in[N]$, $f^{-1}(y)=\{x\colon\,f(x)=y\}$. We use $\occ(y) = |f^{-1}(y)|$.
For a positive integer $k$, $f^{(k)}$ denotes $f$ composed with itself $k$ times, i.e., $f^{(2)}(x)=f(f(x))$.

\begin{definition}
For a function $f \colon [N] \to D$, we denote its \emph{collision probability} by $q(f)$:
\[q(f) = \Pr_{x_1,x_2}[f(x_1) = f(x_2)]\;,\]
where $x_1$ and $x_2$ are sampled independently and uniformly from~$[N]$.
\end{definition}

For a function $f\colon[N]\to[N]$, we will abuse the notation by assuming that $f\colon \F_q\to\F_q$ for a finite field $\F_q$. Indeed, $[N]$ can be embedded into a finite field $\F_q$ with efficient representation and arithmetic for $q\leq 2N$.

A family $\cal{G}$ of functions $g\colon\F_q\to\F_q$ is called \emph{strongly pairwise-independent} if for every $x_1\neq x_2$ and $y_1, y_2$, 
\[
\Pr_{g\sim{\cal{G}}}[g(x_1)=y_1, g(x_2)=y_2] = \frac{1}{q^2}\;.
\]
We will use the strongly pairwise-independent family of hash functions ${\cal{G}}=\{g_{a,b}\colon a,b\in\F_q\}$, where $g_{a,b}(x)=ax+b$. We remark that these functions admit efficient evaluation and succinct representation.

The $\widetilde{O}(\cdot)$ and $\widetilde{\Omega}(\cdot)$ notations suppress factors poly-logarithmic in the input length. 

\subsection{Data Structure Problems}
In this section, we define reporting versions of problems with preprocessing. The difference between the regular (search) version of a problem and its reporting version is that, given a query, the online algorithm outputs \emph{all} solutions to that query. The \FIRep problem is defined as follows.

\begin{definition}\label{def:fi}
The \FIRep problem is solved in two phases by a pair of randomized algorithms $\A=(\A_0, \A_1)$. The algorithms receive oracle access to a function $f\colon[N]\to[N]$, and both $\A_0$ and $\A_1$ can evaluate $f$ at any point $x\in[N]$ in time $\widetilde{O}(1)$.
\begin{description}
\item{\bf Preprocessing phase.} In the first phase, the preprocessing algorithm $\A_0$ preprocesses~$f$ into advice~$\P$ consisting of $S$ bits. 
\item{\bf Query phase.} In the next phase, the online algorithm $\A_1$ receives a query $y \in [N]$ and the advice string~$\P$. If $y \not\in\Im(f)$, then $\A_1$ outputs $\bot$, otherwise $\A_1$ outputs all $x\in f^{-1}(y)$. The running time of the algorithm $\A_1$ is~$T$.
\end{description}
We say that such an algorithm $\A$ for \FIRep is an $(S, T)$-algorithm
if for every function~$f$, with probability at least $1-1/N$ over the randomness of the algorithms, the online algorithm~$\A_1$ correctly answers \emph{all} queries. 
\end{definition}

For an integer $k\geq3$, the problems \kSUMIndRep, \kXORIndRep, \GSIRep, and \JIRep are defined analogously.

\begin{itemize}
    \item \kSUMIndRep: Preprocess a list of~$n$ integers $A=(a_1, \ldots, a_{n}) \in \Z_{\geq0}^n$, and for a query~$y\in\Z_{\geq0}$, output all tuples $(i_1,\ldots,i_{k-1}) \in [n]^{k-1}$ such that $a_{i_1}+\ldots+a_{i_{k-1}}=y$ (output~$\bot$ if no such tuple exists).

    \item \kXORIndRep: Preprocess a list of~$n$ binary vectors $A=(a_1, \ldots, a_{n}) \in (\{0,1\}^\ell)^n$, and for a query~$y\in\{0,1\}^\ell$, output all tuples $(i_1,\ldots,i_{k-1}) \in [n]^{k-1}$ such that $a_{i_1}\oplus\ldots\oplus a_{i_{k-1}}=y$ (output $\bot$ if no such tuple exists).

    \item \GSIRep: Preprocess a string~$S$ of length~$n$. For a query $(P_1,P_2,\alpha,\beta)$, where $P_1$ and $P_2$ are two strings, and $\alpha,\beta\in\Z$, output all pairs $(i, j)$ such that $i+\alpha\leq j\leq i+\beta$, and $P_1$ and $P_2$ occur at positions $i$ and~$j$ of $S$, respectively (output $\bot$ if no such pair exists).
   \item \JIRep: 
    Preprocess a string~$S$ of length~$n$ over an alphabet~$\Sigma$. For a query histogram\footnote{The histogram of a string~$T$ over an alphabet~$\Sigma$ is a vector $h\in\Z_{\geq0}^{|\Sigma|}$, where the $i$th coordinate of $h$ is the number of occurrences of the $i$th character of the alphabet~$\Sigma$ in the string~$T$.} $h\in\Z_{\geq0}^{|\Sigma|}$, output all indices $i$ such that the histogram of a substring of~$S$ starting at the position~$i$ is~$h$ (output $\bot$ if no such~$i$ exists).
\end{itemize}

Next we define problems related to \RangeSearching (see, e.g., \cite{ACDI24}). Let $f \colon [N] \rightarrow [2^w]^d$ and $c : [2^w]^d \rightarrow [2^w]$ be functions computable in $\widetilde{O}(1)$ time, where $w = \polylog(N)$ and $d = O(1)$. The task is to preprocess~$f$ and~$c$ into a data structure that supports the following queries.
\begin{itemize}
    \item \RangeSearching:  Given a $d$-dimensional box $B$, output $x \in [N]$ with $f(x) \in B$, if such an $x$ exists.

    \item \RangeCounting: Given a $d$-dimensional box $B$, output \[\sum _{\substack{x \in[N]\\ f(x) \in B}} c(f(x))\;.\]

    \item \RangeRep: Given a $d$-dimensional box $B$, return all values of $x \in [N]$ such that $f(x) \in B$, or $\bot$ if no such $x$ exists.
\end{itemize}

\section{Data Structure for \texorpdfstring{$\FIRep$}{Function Inversion Reporting}}\label{sec:algorithm}
In this section, we present the proof of our main results for \FIRep. We first restate \cref{thm:basic FIR result}, and show how it implies \cref{prop:BoundedReporting,thm:main}. Then, in \cref{sec:proof}, we present the proof of \cref{thm:basic FIR result}. We remark that the proof of this result is even simpler than the proof of \cite{FN91} for the \FI problem. In particular, for our result, it suffices to use pairwise independent hash functions as opposed to $t$-wise independent hash functions in the result for \FI.

\ThmGeneral*

In the special case where each element in the image has a bounded number of preimages, we immediately recover \cref{prop:BoundedReporting}.
\PropDelta*
\begin{proof}
From the bounded-degree condition $|f^{-1}(y)|\leq\Delta$, we have that the collision probability of~$f$ satisfies
\[q(f) \leq \frac{\Delta}{N}\;.\]
From \cref{thm:basic FIR result}, we obtain the query time $\widetilde{O}(T)$ for
\[
 T = \frac{N^3}{S^2} \cdot q(f) + \frac{N}{S} \cdot \occ(y)
 \leq\frac{N^3}{S^2} \cdot  \frac{\Delta}{N} + \frac{N}{S} \cdot \Delta
\leq \frac{2N^2\cdot \Delta}{S^2}\;.\qedhere
\]
\end{proof}

Finally, we are ready to present the proof of~\cref{thm:main}.
\ThmMain*
\begin{proof}
Let $p_0=\ceil{\log(N/S)}$, and $p_1=\ceil{\log(N)}$. We will use data structures $D_{p_0},\ldots,D_{p_1}$, where for $p_0\leq p\leq p_1$, the data structure $D_p$ outputs all preimages of $y$ for every $y$ satisfying $\occ(y)\leq2^p$.
For $p_0\leq p\leq p_1$, we define the set of heavy images as follows,
\[
H_p = \{y \in [N]\colon \occ(y) >2^p \} \;.
\]
Note that $H_{p_0} \supseteq \ldots\supseteq H_{p_1}$.
The data structure $D_{p_0}$ will store all $y\in H_{p_0}$ along with $\occ(y)$ in space $\widetilde{O}(S)$.

For every $p_0\leq p\leq p_1$, we define the function $f_p\colon[2N]\to[2N]$ as follows,
\begin{align}\label{eq:fp}
f_p(x) = 
\begin{cases}
f(x), &\text{ if } x\in[N] \text{ and } f(x)\not\in H_p\;,\\
x+N, &\text{ if } x\in[N] \text{ and } f(x)\in H_p\;,\\
x, &\text{ if } x\not\in[N]\;.
\end{cases}
\end{align}
Note that by the definition of the function $f_p$, we have that for every $y\in[2N]$, 
\[
|f_p^{-1}(y)|\leq \max(2^p, 2)\leq 2^{p+1}\;.
\]
\paragraph{Preprocessing algorithm.}
For every $p_0\leq p\leq p_1$, we create a data structure $D_p$ from \cref{prop:BoundedReporting} using space $S$. This results in preprocessing time $\widetilde{O}(N)$ and space complexity $\widetilde{O}(S)$.

\paragraph{Online algorithm.}
For a query $y\in[N]$, we first find the value of $p$ such that the data structure $D_p$ allows us to compute $f^{-1}(y)$,
\[
p =
\begin{cases}
    p_0, \text{ if } y \not\in H_{p_0}\;,\\
    \ceil{\log(\occ(y))}, \text{ if } y \in H_{p_0}\;.
\end{cases}
\]
Note that in the case $y \in H_{p_0}$, the value $\occ(y)$ is stored in the preprocessing stage, and, thus, $p$ can be computed efficiently.
We have that $y\not \in H_p$, and, thus, $f^{-1}(y)=f_p^{-1}(y)$. The online algorithm uses the data structure $D_p$ to compute $f_p^{-1}(y)$. Note that $f_p$ is efficiently computable for all $p$, as $D_{p_0}$ stores the relevant $\occ$ information for all inputs.
By the guarantees of \cref{prop:BoundedReporting}, the running time of this step is $\widetilde{O}(T)$ for 
\[
T = \frac{(2N)^2}{S^2}\cdot\Delta
\leq \frac{(2N)^2}{S^2}\cdot2^{p+1}
\leq \frac{(2N)^2}{S^2}\cdot 4(N/S + \occ(y))
= O\left(\frac{N^3}{S^3}+\frac{N^2}{S^2}\cdot\occ(y)\right) 
\;,
\]
where we used that 
\[2^{p+1}\leq 2\left(2^{p_0}+2^{\ceil{\log(\occ(y))}}\right)\leq 2\left(2N/S+2\cdot\occ(y)\right)\leq 4\left(N/S+\occ(y)\right)\;.\qedhere\]
\end{proof}
\subsection{Proof of \texorpdfstring{\cref{thm:basic FIR result}}{Theorem~\ref{thm:basic FIR result}}}\label{sec:proof}
\begin{proof}[Proof of \cref{thm:basic FIR result}]
For positive integer parameters $m, \ell, t$, the data structure works with {\bf $\ell$ tables, each containing $m$ chains of length $t$}. We set the parameters as follows. 
\begin{align*}
t &= N/S \;,\\
m &= \frac{S}{3Nq(f)}\;,\\
\ell &= 36 q(f)\cdot N\log{N}\;.
\end{align*}
If $q(f) > \frac{S^2}{3N^2}$, then $T> N/3$, and we can solve the problem trivially in time $\widetilde{O}(T)$. Thus, in the following we assume that  $q(f)\leq \frac{S^2}{3N^2}$. Under this assumption we have that $m\geq 1$.
We will use the following bounds:
\begin{align}
    mtq(f) &\leq 1/3 \label{eq:mtq}\;,\\
    12N\log{N} &\leq mt\ell = \widetilde{O}(N) \label{eq:mtl}\;,\\
    m\ell +t\ell&= \widetilde{O}(S)\label{eq:ml}\;,\\
    mt^2\ell \left(tq(f)+\frac{\occ(y)}{N}\right) &= \widetilde{O}\left(\frac{N^3}{S^2}\cdot q(f)+\frac{N}{S}\cdot\occ(y)\right)\label{eq:runtime}\;,\\
    t^2\ell&=\widetilde{O}\left(\frac{N^3}{S^2}\cdot q(f)\right)\label{eq:runtimecheck}\;.
\end{align}

\paragraph{Preprocessing algorithm.} In the preprocessing phase we first compute $q(f)$ in time~$\widetilde{O}(N)$ and use it to compute the parameters $t,m,\ell$.
Then, for each $i\in[\ell]$ and each $0\leq a\leq t-1$, we independently sample $g_{i,a}\colon[N]\to[N]$ from a family of strongly pairwise-independent (and efficiently computable) functions. For each $i\in[\ell]$, we also independently and uniformly at random choose $m$ roots of chains $X_0^{i,1},\ldots,X_0^{i,m}\in[N]$. For each $i\in[\ell]$ and $j\in[m]$, we define 
\[
X_{a+1}^{i,j}=g_{i,a}(f(X_a^{i,j}))\;,\;\;\; 0\leq a\leq t-1 \;.
\]
We define the length-$t$ chain $C^{i,j}=\{X_0^{i,j},\ldots,X_{t-1}^{i,j}\}$, and call $X_t^{i,j}$ its stored endpoint. For $i\in[\ell]$, the table $T_i$ stores the pairs $(X^{i,j}_t, X^{i,j}_0)$ for all $j\in[m]$. The data structure stores all tables $T_i$. Let the leftover domain elements not covered by any chain be
\[
L := [N] \setminus \{X_a^{i,j}\colon i\in[\ell], j\in[m], 0\leq a\leq t-1   \} \;.
\]
If $L\neq\emptyset$, then we abort. 

\paragraph{Preprocessing time and success probability.} The running time of the preprocessing routine is $\widetilde{O}(\ell m t)=\widetilde{O}(N)$ (\cref{eq:mtl}). We now show that the preprocessing stage aborts with probability at most~$1/N^2$. Fix $i\in[\ell]$ and $j\in[m]$, and consider the chain $C^{i,j}$. All $t$ elements in $C^{i,j}$ are independent and uniform in $[N]$ (indeed, conditioned on the prefix of the chain, the next chain element uses a fresh hash function). Therefore, for every $x\in[N]$,
\[
\Pr[x \in C^{i,j}] = 1 - (1-1/N)^t \geq \frac{t}{2N} \;,
\]
where the inequality uses $t\leq N$. Next we show that for $j\neq j'$,
\[
\Pr[x \in C^{i,j} \text{ and } x \in C^{i,j'}] \leq \frac{3t^2 q(f)}{2N} \;.
\]
For $a\neq b$, $X^{i,j}_a$ and $X^{i,j'}_b$ are independent and uniform. (Indeed, for example, if $a<b$, then $X^{i,j'}_b$ is obtained by applying a hash function independent of $X^{i,j}_a$.) By taking the union bound over all $a\neq b$, we have that in this case, the probability is bounded from above by $\frac{t(t-1)}{N^2}$.

For $a=b$, consider the first index $u$ such that $f(X^{i,j}_u)=f(X^{i,j'}_u)$ (and $u=\infty$ if no such index exists). Let $E$ be the event that $u<a$. Since $X^{i,j}_a=X^{i,j'}_a$ implies that $u\leq a$, we have that
\[
\Pr[X^{i,j}_a=X^{i,j'}_a=x] \leq \Pr[E \text{ and } X^{i,j}_a=X^{i,j'}_a=x] + \Pr[\neg{E} \text{ and } X^{i,j}_a=X^{i,j'}_a=x] \;.
\]

Conditioned on $\neg{E}$, we have that $X^{i,j}_a$ and $X^{i,j'}_a$ are independent and uniform, giving an upper bound of $\Pr[\neg{E} \text{ and } X^{i,j}_a=X^{i,j'}_a=x]\leq \frac{1}{N^2}$. For the other case, the probability that $u<a$ is at most $a\cdot q(f)$ by the bound $q(f)$ on the collision probability of~$f$ and the union bound over all $u<a$. Conditioned on the event~$E$, we have that $X^{i,j}_a=X^{i,j'}_a$, and that the value of $X^{i,j}_a=X^{i,j'}_a$ is uniformly random in $[N]$. Therefore, $\Pr[E \text{ and } X^{i,j}_a=X^{i,j'}_a=x]\leq\frac{a\cdot q(f)}{N}$. Taking the union bound over all $0\leq a\leq t-1$, and combining it with the bound for $a\neq b$, we conclude that
\[
\Pr[x \in C^{i,j} \text{ and } x \in C^{i,j'}] \leq \frac{t^2}{N^2}+\frac{t^2 q(f)}{2N} \leq \frac{3t^2 q(f)}{2N}\;,
\]
where the last inequality uses $q(f)\geq 1/N$.

We now use the two inequalities above along with the inclusion-exclusion bound to bound the probability that a fixed $x$ appears in one of the $m$ chains of a fixed table. Fix an $i\in[\ell]$ and $x\in[N]$.
\begin{align*}
\Pr[x \in \bigcup_{j\in[m]} C^{i,j}] &\geq 
\sum_{j\in[m]} \Pr[x \in C^{i,j}] - \sum_{1\leq j<j'\leq m} \Pr[x \in C^{i,j} \text{ and } x\in C^{i, j'}]\\
&\geq \frac{mt}{2N} - \frac{3m^2t^2q(f)}{4N}\\
&\geq \frac{mt}{4N}\;,
\end{align*}
where the last inequality uses $mtq(f)\leq 1/3$ (\cref{eq:mtq}).

Finally, since the $\ell$ tables are independent, the probability that a fixed $x$ is not covered by any chain in any of the $\ell$ tables is
\[
\Pr[x \in L] \leq \left(1-\frac{mt}{4N}\right)^{\ell} \leq e^{\frac{-mt\ell}{4N}}
\leq \frac{1}{N^3}
\;,
\]
where the last inequality uses $mt\ell\geq 12N \ln{N}$ (\cref{eq:mtl}). Therefore $\E[\,|L|\,]\leq\frac{1}{N^2}$, and, by Markov's inequality, $\Pr[|L|\geq1]\leq \frac{1}{N^2}$.

\paragraph{Space complexity.}
The data structure stores $\ell$ tables, each of which stores $\widetilde{O}(m)$ elements, resulting in space complexity of $\widetilde{O}(\ell m)$. Storing $\ell t$ hash functions takes space $\widetilde{O}(\ell t)$. Therefore, the total space complexity of the data structure is $\widetilde{O}(\ell m+\ell t)=\widetilde{O}(S)$ (\cref{eq:ml}).

\paragraph{Online algorithm.}
Let $y\in[N]$ be the query to be inverted with $\occ(y)=|f^{-1}(y)|$. For each table $i\in[\ell]$ and each potential position in a chain $0\leq a\leq t-1$, we compute the continuation of the chain:
\begin{align*}
W_{a+1} &= g_{i,a}(y)\;,\\ 
W_{b+1} &= g_{i,b}(f(W_b)) \text{ for } a+1\leq b\leq t-1 \;.
\end{align*}
Next, we go over all pairs in $T_i$ whose first component is $W_t$. For each such pair $(X_t^{i,j}, X_0^{i,j}) \in T_i$ with $X_t^{i,j}=W_t$, we compute $X_a^{i,j}$ in time $\widetilde{O}(t)$. We report $X_a^{i,j}$ if $f(X_a^{i,j})=y$ and it has not yet been reported.

Now we argue the correctness of the online algorithm. The check $f(X_a^{i,j})=y$ guarantees that the algorithm does not report false positives. Since $L=\emptyset$, each $x\in[N]$ lies in at least one chain $C^{i,j}$. Then if $x=X^{i,j}_a$, and $y=f(x)$, then $W_{a+1}=g_{i,a}(y)=X^{i,j}_{a+1}$, and $W_t=X^{i,j}_t$. Thus, every preimage of $y$ is covered by a chain and is reported by the online algorithm.

\paragraph{Query time.} The time to compute continuations of chains for all $i\in[\ell]$ and $0\leq a\leq t-1$ is 
\begin{align}
    \widetilde{O}(t^2\ell) \;.\label{eq:checktime}
\end{align}

Next we bound the number of times a match is found by the online algorithm, that is, $X_t^{i,j}=W_t$. Let us fix $y$, the table index $i\in[\ell]$, the chain index $j\in[m]$, and the index of the guessed position $0\leq a \leq t-1$. We analyze the probability that $X_t^{i,j}=W_t$ over the randomness of the preprocessing phase. 

$X_a^{i,j}$ is uniformly random in $[N]$, and, thus, $\Pr[f(X_a^{i,j})=y]=\frac{\occ(y)}{N}$. 
For $u\geq a+1$, let $E_u$ be the event that $f(X_u^{i,j})= f(W_u)$, and let $E_a$ be the event that $f(X_a^{i,j})=y$. The event $\neg{E_{u-1}}$ implies that $X_{u}$ and $W_u$ are independent and uniformly random, and, thus, $\Pr[E_u|\neg{E_{u-1}}]\leq q(f)$. From this we conclude that
\begin{align*}
\Pr[X_t^{i,j}=W_t] &\leq \Pr[\bigcup_{u\geq a}E_u]\\
&\leq \Pr[E_a] + \sum_{u>a} \Pr[E_u | \neg{E_a},\ldots,\neg{E_{u-1}}]\\
&=\Pr[E_a] + \sum_{u>a} \Pr[E_u | \neg{E_{u-1}}]\\
&\leq \frac{\occ(y)}{N}+ (t-a)q(f) \;.
\end{align*}
By linearity of expectation, the expected number of matches found by the algorithm is at most
\begin{align*}
\ell m \sum_{a=0}^{t-1}\left(\frac{\occ(y)}{N}+(t-a)q(f)\right) = O\left(\ell m t \left(\frac{\occ(y)}{N}+tq(f)\right)\right)\;.
\end{align*}
Each such match takes time $\widetilde{O}(t)$ to reconstruct a chain of length~$t$. Together with \cref{eq:checktime}, we have that the expected running time of the algorithm for the query $y$ is
\begin{align*}
\widetilde{O}\left(t^2\ell +\ell m t^2\left(\frac{\occ(y)}{N}+tq(f)\right)\right)
= \widetilde{O}\left(\frac{N^3}{S^2}\cdot q(f)+\frac{N}{S}\cdot\occ(y)\right)\;,
\end{align*}
where the last bound uses \cref{eq:runtime,eq:runtimecheck}.

\paragraph{Success probability amplification.}
We have that for a fixed $y$, the expected running time of the online algorithm is bounded from above by $\widetilde{T}=C\log^c(N)\left(\frac{N^3}{S^2}\cdot q(f)+\frac{N}{S}\cdot\occ(y)\right)$ for absolute constants $c, C>0$. We run the online algorithm for $2\widetilde{T}$ steps, and abort it if it is not finished. By Markov's inequality, it succeeds with probability at least $1/2$. Now we build $3\log{N}$ independent copies of the data structure. Then each $y$ is successfully inverted in time $2\widetilde{T}$  with probability at least $1-\frac{1}{N^3}$. By the union bound, all $y$ are successfully inverted with probability at least $1-\frac{1}{N^2}$. In the case where $\occ(y)$ is not known (and, thus, $\widetilde{T}$ is not known), we run the $3\log{N}$ copies of the online algorithm in parallel, and return the output of the first copy that finishes. The amplification step incurs $\widetilde{O}(1)$ increase in time and space complexities that are hidden in the $\widetilde{O}$ notation. This finishes the proof of the theorem.
\end{proof}

\section{Extensions}\label{sec:extensions}
In this section, we show that the proof of \cref{thm:main} extends to the case of functions with larger ranges, to the counting version of \FI, and to the setting where the preprocessing and online algorithms share randomness (or are non-uniform).

\subsection{\FIRep with Arbitrary Ranges}
Corrigan-Gibbs and Kogan showed that the complexity of \FI does not change if the size of the range of the function $f$ grows polynomially~\cite[Lemma~21]{CK19}. In this section, we remark that this holds true for \FIRep as well. Indeed, for a function $f\colon[N]\to[N^c]$, we can use universal hashing to reduce the domain from $[N^c]$ to $[N]$ as in~\cite[Lemma~21]{CK19}, and for at least one of the $O(c\log{N})$ hashing functions, the collision probability does not change asymptotically, which is sufficient for extending the results of \cref{thm:main} and \cref{thm:basic FIR result} to the case of functions with polynomially large ranges.

Alternatively, one can directly modify the proofs of \cref{thm:main} and \cref{thm:basic FIR result} to make them work for functions $f\colon[N]\to[N^c]$, as we explain below.
\begin{theorem}\label{thm:4withlarge}
 For every $1\leq S \leq N$ and every constant $c\geq1$, there exists a data structure that preprocesses an arbitrary function $f \colon [N] \to [N^c]$ into space~$S$, and for any query $y \in [N^c]$, it outputs all $\occ(y)$ preimages of $y$ in time~$\widetilde{O}(T)$ for 
    \[
    T = \frac{N^3}{S^2} \cdot q(f) + \frac{N}{S} \cdot \occ(y)\;.
    \]
    With probability $1-1/N$ over the randomness of the preprocessing stage, the data structure answers all queries correctly, and the preprocessing time is $\widetilde{O}(N)$.
\end{theorem}
The proof of \cref{thm:4withlarge} differs from the proof of \cref{thm:basic FIR result} only in that (i) $y$ now comes from~$[N^c]$, (ii) the hash functions $g_{i,a}$ map $[N^c]$ to $[N]$, (iii) the seed lengths of hash functions change by a constant factor, and (iv) the probability amplification procedure runs $O(c\log{N})$ independent copies of the data structure instead of $3\log{N}$ copies.
\begin{theorem}\label{thm:3withlarge}
    For every $1\leq S\leq N$ and every constant $c\geq1$, there exists a data structure that preprocesses an arbitrary function $f \colon [N] \to [N^c]$ into space~$\widetilde{O}(S)$, and for any query $y \in [N^c]$, it outputs all $\occ(y)$ preimages of $y$ in time~$\widetilde{O}(T)$ for 
    \[
    T = \frac{N^3}{S^3} + \frac{N^2}{S^2}\cdot \occ(y)\;.
    \]
With probability $1-1/N$ over the randomness of the preprocessing stage, the data structure answers all queries correctly, and the preprocessing time is $\widetilde{O}(N)$.
\end{theorem}
The proof of \cref{thm:3withlarge} follows the proof of \cref{thm:main}, replacing $f_p$ from \cref{eq:fp} by $f_p\colon[N]\to[N^c+N]$ defined as
\begin{align*}\label{eq:fplarge}
f_p(x) = 
\begin{cases}
f(x), &\text{ if } f(x)\not\in H_p\;,\\
x+N^c, &\text{ if } f(x)\in H_p\;.\\
\end{cases}
\end{align*}
Then \cref{thm:3withlarge} simply calls \cref{thm:4withlarge} instead of \cref{thm:basic FIR result}.

\subsection{\FICount}
We show that the proof of \cref{thm:main} with minor modifications gives a data structure that in ``Function-Inversion-time'' solves the (weighted) counting version of \FI! In the following, we will assume that both the preprocessing and the online algorithms have access to efficiently computable functions $f$ and $w$. Furthermore, we will assume that the values of $w$ and their sums can be stored in a unit of space.
\begin{theorem}\label{thm:counting}
 For every $1\leq S \leq N$ and every constant $c$, there exists a data structure that preprocesses arbitrary functions $f\colon [N] \to [N^c]$ and $w\colon [N]\to\R$ into space~$\widetilde{O}(S)$, and for any query $y \in [N^c]$, it outputs the weighted sum $\sum_{x\colon f(x) = y} w(x)$ in time $\widetilde{O}(T)$ for
 \[
 T = \frac{N^3}{S^3}
 \;.
 \] 
With probability $1-1/N$ over the randomness of the preprocessing stage, the data structure answers all queries correctly, and the preprocessing time is $\widetilde{O}(N)$.
\end{theorem}
\begin{proof}
   We define the set of heavy images as follows,
\[
H = \{y \in [N]\colon \occ(y) >N/S \} \;.
\]
Since $|H|< S$, the data structure stores the (sorted) set of all $y\in H$ along with the answer to the query $y$, in space $\widetilde{O}(S)$.

Next, we simply use the data structure of \cref{thm:3withlarge} for the function $f$ with preprocessing space~$S$. Given a query $y$, the online algorithm first checks if $y\in H$, in which case it returns the stored answer for the query. Otherwise, it runs the online algorithm of the data structure from \cref{thm:3withlarge} for the reporting version of the problem, and outputs the sum of the corresponding values of $w$.

By the guarantees of \cref{thm:3withlarge}, the query time is $\widetilde{O}(T)$ for
\[
T = \max_{y\not\in H} \left(\frac{N^3}{S^3} + \frac{N^2}{S^2}\cdot \occ(y)\right)\leq
\frac{N^3}{S^3} + \frac{N^2}{S^2}\cdot \frac{N}{S}
=O\left(\frac{N^3}{S^3}\right)\;. \qedhere
\]
\end{proof}

\subsection{\FIRep with Shared Randomness}
\cite{GGPS23} observed that if the preprocessing and online algorithms are non-uniform (or share randomness), then the data structure of \cite{FN91} can be used to achieve the tradeoff of $ST\approx N^{3/2}$, which is better than the classical tradeoff for $T\gg S$. In this section, we generalize this result to the case of \FIRep. 

We present the result below in the model with shared randomness, where both the online and preprocessing algorithms have access to a shared random string. We remark that one can remove this assumption at the expense of making the algorithms non-uniform~\cite[Section~6]{GGPS23}. Another drawback of the online algorithms in~\cite{GGPS23} and in \cref{thm:sharedrandomness} is that they use space~$T$, which may exceed the space $S$ used by the data structure in the preprocessing stage. Finally, we note that in applications of such \FI results to other data structure problems, both drawbacks (shared randomness, and space complexity of the online algorithms) can often be removed (see~\cite{DG26}).

\begin{theorem}\label{thm:sharedrandomness}
   For every $1\leq S\leq N$, there exists a data structure with shared randomness that preprocesses an arbitrary function $f \colon [N] \to [N]$ into space~$\widetilde{O}(S)$, and for any query $y \in [N]$, it outputs all $\occ(y)$ preimages of $y$ in time~$\widetilde{O}(T)$ for 
    \[
    T = \frac{N^{3/2}}{S} + \frac{N^2}{S^2}\cdot \occ(y)\;.
    \]
With probability $1-1/N$ over the randomness of the preprocessing stage and the shared random string, the data structure answers all queries correctly, and the preprocessing time is $\widetilde{O}(N)$.
\end{theorem}
\begin{proof}
If $S=\widetilde{O}(\sqrt{N})$, then $T=\widetilde{\Omega}(N)$, and the problem can be solved trivially. Thus, in the following we will assume that $S=\widetilde{\Omega}(\sqrt{N})$.

The proof is almost identical to the proof of \cref{thm:main} with the main difference being the idea from~\cite{GGPS23}: the preprocessing algorithm does not store the lists of heavy elements $H_p$, but instead both the preprocessing and the online algorithms use the shared randomness to sample these lists.

Specifically, we set $M=\Theta(N^{3/2}\log(N)/S)$, and sample $M$ points $x_1,\ldots,x_M \in [N]$ using shared randomness. From this, we compute the values $y_1=f(x_1),\ldots,y_M=f(x_M)$. For each such $y$, let $c(y)=|\{i\colon f(x_i)=y\}|$. 
We set $p_0 = \ceil{\log(S/\sqrt{N})}$ and $p_1=\ceil{\log(N)}$.
For every $p_0\leq p\leq p_1$, both the preprocessing and the online algorithms recover the same sets $H'_p$ without storing them in the data structure, where
\[
H'_p= \left\{y\in[N]\colon c(y) \geq \frac{2M2^p}{N} \right\} \;.
\]
By Chernoff's bound, with probability $1-1/N^{2}$, for every $y$ and every $p_0\leq p\leq p_1$, we have that
\begin{align*}
    y \in H'_p  \quad &\implies \quad \occ(y)\geq 2^p\;,\\
    \occ(y) \geq 2^{p+2} \quad &\implies \quad y \in H'_p \;.
\end{align*}
For each $p_0\leq p\leq p_1$, the function $f_p$ is defined analogously to \cref{eq:fp} using $H'_p$ in place of~$H_p$. Then \cref{prop:BoundedReporting} gives us the query time $\widetilde{O}(T)$ for
\begin{align*}
T &= M+\frac{(2N)^2}{S^2}\cdot\Delta \\
&= \widetilde{O}\left(\frac{N^{3/2}}{S}+\frac{N^2}{S^2}\cdot 2^p\right)\\
&= \widetilde{O}\left(\frac{N^{3/2}}{S}+\frac{N^2}{S^2}\cdot\left(\frac{S}{\sqrt{N}}+\occ(y)\right)\right)\\
&=\widetilde{O}
\left(\frac{N^{3/2}}{S} + \frac{N^2}{S^2}\cdot \occ(y)\right)
\;.\qedhere
\end{align*}

\end{proof}

\section{Applications}\label{sec:apps}
In this section, we demonstrate how a simple application of \cref{thm:main} improves the state-of-the-art data structures for several problems.
\subsection{\texorpdfstring{$\kSUMInd$}{kSum-Indexing}}\label{sec:3sum}
\cite{BGLPRS24} showed a reduction from \ThreeSUMIndRep to \ThreeSUMInd. The reduction implies that if there is an $(S, T)$-data structure for \ThreeSUMInd, then there is an $(S, T(1+\occ(y)))$-data structure for \ThreeSUMIndRep. Together with the result of \cite{GGHPV20}, this reduction implies that \ThreeSUMIndRep can be solved with space $S=n^{2-\delta}$ and time $T=n^{3\delta}(1+\occ(y))$. Applying the same reduction to the result of~\cite{DG26}, we obtain a data structure for \ThreeSUMIndRep with $S=n^{2-\delta}$ and $T=n^{1/2+\delta}(1+\occ(y))$, which is better than the previous one for $\delta\in(1/4,1/2)$. Thus, the best known data structure for \ThreeSUMIndRep has space complexity $S=n^{2-\delta}$ and query time $\widetilde{O}(T)$ for
\[
T = \min\left(n^{3\delta}+n^{3\delta}\cdot\occ(y)\;,\; n^{1/2+\delta}+n^{1/2+\delta}\cdot\occ(y)  \;,\;
n\right)\;.
\]
In the next lemma, we show that an immediate application of \cref{thm:main} improves both of these tradeoffs. We remark that the second tradeoff $n^{1/2+\delta}+n^{2\delta}\cdot\occ(y)$ in the lemma below  is better than $n^{1/2+\delta}+n^{1/2+\delta}\cdot\occ(y)$ in the whole range $\delta\in(1/4,1/2)$ where this data structure outperforms the first data structure and the trivial data structure with $T=O(n+\occ(y))$.

\begin{lemma}\label{lem:3sum}
For every $0\leq\delta\leq1$, there exists a data structure that solves \ThreeSUMIndRep using space $S=\widetilde{O}(n^{2-\delta})$ and query time $\widetilde{O}(T)$ for
\[
T=n^{2\delta}\cdot\occ(y)+\min\left(
n^{3\delta}
\;,\;
n^{1/2+\delta}
\right)
\;.
\]
\end{lemma}
\begin{proof}
We use \cref{thm:main} with the reduction of~\cite{GGHPV20} for the first tradeoff, and \cref{thm:sharedrandomness} with the reduction of~\cite{DG26} for the second tradeoff.

Let $(a_1,\ldots,a_n)$ be the input of \ThreeSUMIndRep. By \cite[Theorem~7]{GGHPV20}, we can without loss of generality assume that $a_i=\widetilde{O}(n^2)$ for all $i\in[n]$. Let us now define the function $f(i,j)=a_i+a_j$ for $i,j\in[n]$. We note that in order to answer a \ThreeSUMIndRep query $y\in\Z$, it is sufficient to solve the \FIRep problem for the function $f$ and the query point $y$. Since the function $f$ is defined with the domain and range of size $N=\widetilde{O}(n^2)$, \cref{thm:main} gives us the first tradeoff.

Let $p, q = \widetilde{\Theta}(n)$ be two random primes. From the Chinese Remainder Theorem, it is sufficient to find all $a_i$ and $a_j$ satisfying $a_i + a_j = y \bmod p$ and $a_i + a_j = y \bmod q$.

For every $0\leq r\leq q-1$ and $i \in [n]$, we define the function $h_r(i)$ as follows,
\[
h_r(i)=
\begin{cases}
a_i+a_j \bmod p, & \text{ for the smallest } j \text{ s.t. } a_i+a_j= r\bmod q \text{ if it exists}\;,\\
\text{random}, & \text{ otherwise}\;.
\end{cases}
\]

In the preprocessing stage, we create the data structures from \cref{thm:sharedrandomness} for each of the $q=\widetilde{O}(n)$ functions $h_r$. In the query phase, we compute $h^{-1}_{y \bmod q}(y \bmod p)$. We refer the reader to \cite[Theorem~5.1]{DG26} for the analysis showing that with high probability this gives all solutions to the given instance of \ThreeSUMIndRep. Now the guarantee of \cref{thm:sharedrandomness} gives us the second tradeoff.
\end{proof}

Similarly, we can obtain the corresponding improvements for the related problems of \kSUMIndRep and \kXORIndRep. The best known data structures~\cite{GGHPV20,DG26} have space $S=n^{k-1-\delta}$ and query time $\widetilde{O}(T)$ for
\[
T = \min\left( n^{3\delta}+n^{3\delta}\cdot\occ(y) 
\;,\;
n^{1/2+\delta}+n^{1/2+\delta}\cdot\occ(y)
\right)\;,
\]
where the second tradeoff is better for $\delta\in(1/4,1/2)$.
\begin{lemma}
For every constant $k\geq 3$ and every $0\leq\delta\leq k-2$, there exists a data structure that solves \kSUMIndRep and \kXORIndRep using space $S=\widetilde{O}(n^{k-1-\delta})$ and query time $\widetilde{O}(T)$ for \[
T=n^{2\delta}\cdot\occ(y)+\min\left(
n^{3\delta}
\;,\;
n^{1/2+\delta}
\right)
\;.
\]
\end{lemma}
\begin{proof}
The result follows by a reduction from \kSUMIndRep to \ThreeSUMIndRep on a set of size $O(n^{k-2})$ of all $(k-2)$-wise sums of the inputs and the input set of size~$n$, and an analogue of \cref{lem:3sum} for unbalanced sets. Similarly, the result for \kXORIndRep follows by such a reduction and an analogue of \cref{lem:3sum} for \ThreeXORIndRep. See \cite[Theorem~2]{DG26} for the rigorous proof of the reduction.
\end{proof}

\subsection{Problems on Strings}
\cite{BGLPRS24} gave beautiful reductions from \JIRep and \GSIRep to \ThreeSUMIndRep. As an immediate consequence of these reductions and \cref{lem:3sum}, we obtain improved data structures for these problems.

\begin{lemma}
    For every $0 \leq \delta \leq 1$, there exists a data structure that solves \JIRep using space $S = \widetilde{O}(n^{2 - \delta})$ and query time $\widetilde{O}(T)$ for
\[ 
T=n^{2\delta}\cdot\occ+\min\left(
n^{3\delta}
\;,\;
n^{1/2+\delta}
\right)
\;.
\]
\end{lemma}
\begin{proof}
This lemma follows from \cref{lem:3sum} and the reduction from~\cite[Corollary~2]{BGLPRS24}.
\end{proof}

\begin{lemma}
    For every $0 \leq \delta \leq 1$, there exists a data structure that solves \GSIRep using space $S = \widetilde{O}(n^{2 - \delta})$ and query time $\widetilde{O}(T)$ for
\[ 
T=n^{2\delta}\cdot\occ+|P_1|+|P_2|+\min\left(
n^{3\delta}
\;,\;
n^{1/2+\delta}
\right)
\;.
\]
\end{lemma}
\begin{proof}
This lemma follows from \cref{lem:3sum} and the reduction from~\cite[Theorems~6 and~7]{BGLPRS24}.
\end{proof}

\subsection{Range Reporting}
Recently \cite{ACDI24} used \FI to solve \RangeRep and \RangeCounting problems. Let $B$ be a query box in the \RangeRep problem, and let $\occ(B)$ denote the number of points $x\in[N]$ s.t. $f(x)\in B$. The best known data structures~\cite{ACDI24} for \RangeRep and \RangeCounting use space $S=N^{1-\delta}$ and query time~$\widetilde{O}(T)$ for 
\begin{align*}
T &= N^{3\delta} + N^{3\delta}\cdot \occ(B)\;,\text{and}\\
T &= N^{4\delta} \;,
\end{align*}
respectively. 
Below we show that \cref{thm:counting} allows us to solve \RangeCounting in the ``Range-Search-time'', and \cref{thm:main} gives us an improved data structure for \RangeRep.

\begin{lemma}
For every $0\leq\delta\leq1$, there exists a data structure that solves \RangeRep using space $S=\widetilde{O}(N^{1-\delta})$ and query time $\widetilde{O}(T)$ for
\[
T=N^{3\delta}+N^{2\delta}\cdot \occ(B)
\;,
\]
where $\occ(B)$ denotes the number of points $x\in[N]$ s.t. $f(x)\in B$ for the query box~$B$.
\end{lemma}
\begin{proof}
    The lemma follows from \cref{thm:main} and the reduction from \cite[Theorem~9]{ACDI24}.
\end{proof}

\begin{lemma}
For every $0\leq\delta\leq1$, there exists a data structure that solves \RangeCounting using space $S=\widetilde{O}(N^{1-\delta})$ and query time $\widetilde{O}(T)$ for
\[
T=N^{3\delta}
\;.
\]    
\end{lemma}
\begin{proof}
     This lemma follows from \cref{thm:counting} and the reduction from \cite[Theorem~4]{ACDI24}.
\end{proof}

\subsection{Nearest Neighbors}
As our final application, we observe that we can slightly improve the tradeoff for the \ANN problem from~\cite{M24}. Indeed, \cite{M24} designs a data structure for the All-Function Inversion problem for locality-sensitive hash functions. We show that \cref{thm:main} improves the tradeoff for this problem, too. The best known data structure for this problem (\cite[Theorem~6]{M24} and \cite[Theorem~10]{ACDI24}) uses space $S=\widetilde{O}(N+RN^{1-\delta})$ and query time $\widetilde{O}(T)$ for
\[
T = t N^{3\delta} + t N^{3\delta} \cdot |f_i^{-1}(y)|\;.
\]

\begin{theorem} \label{LSH}
    Let $f_1, \ldots, f_R\colon [N] \to D$ be functions each computable in time $t$. For every $0\leq\delta\leq1$, there is a randomized data structure with preprocessing time $\widetilde{O}(tRN)$ that uses space $S = \widetilde{O}(N + RN^{1-\delta})$, and for every query $(i,y)$, it reports $f_i^{-1}(y)$ in time $\widetilde{T}$ for
    \[
T = t N^{3\delta} + t N^{2\delta} \cdot |f_i^{-1}(y)|\;.
\]
\end{theorem}
\begin{proof}
This result follows by applying the reporting data structure from \cref{thm:main} to each of the~$R$ functions $f_1,\ldots,f_R$.
\end{proof}

\subsection*{Acknowledgments}
We are thankful to anonymous reviewers for their helpful suggestions.

\subsection*{AI Disclosure}
An LLM was used exclusively for language editing in a few specific places in the paper. The results are proven and written by the authors.

\clearpage
\bibliographystyle{alpha}
\bibliography{refs} 

@inproceedings{Oechslin03,
  author       = {Oechslin, Philippe},
  title        = {Making a Faster Cryptanalytic Time-Memory Trade-Off},
  booktitle={CRYPTO},
  year         = {2003},
}

@article{H80,
  title={A cryptanalytic time-memory trade-off},
  author={Hellman, Martin},
  journal={IEEE Transactions on Information Theory},
  volume={26},
  number={4},
  pages={401--406},
  year={1980},
  publisher={IEEE}
}

@inproceedings{BGLPRS24,
  title={Gapped String Indexing in Subquadratic Space and Sublinear Query Time},
  author={Bille, Philip and G{\o}rtz, Inge Li and Lewenstein, Moshe and Pissis, Solon P. and Rotenberg, Eva and Steiner, Teresa Anna},
  booktitle={STACS},
  year={2024},
}

@inproceedings{GGHPV20,
  title = {Data Structures Meet Cryptography: {{3SUM}} with Preprocessing},
  booktitle = {{STOC}},
  author = {Golovnev, Alexander and Guo, Siyao and Horel, Thibaut and Park, Sunoo and Vaikuntanathan, Vinod},
  year = {2020},
}

@inproceedings{ACDI24,
  title={A General Technique for Searching in Implicit Sets via Function Inversion},
  author={Aronov, Boris and Cardinal, Jean and Dallant, Justin and Iacono, John},
  booktitle={SOSA},
  year={2024},
}

@inproceedings{FN91,
author = {Fiat, Amos and Naor, Moni},
title = {Rigorous time/space tradeoffs for inverting functions},
year = {1991},
booktitle = {STOC},
}

@inproceedings{M24,
  title={Improved Space-Efficient Approximate Nearest Neighbor Search Using Function Inversion},
  author={McCauley, Samuel},
  booktitle={ESA},
  year={2024},
}

@inproceedings{BBM26,
  title={Space-efficient text indexing with mismatches using function inversion},
  author={Bibbens, Jackson and Borevitz, Levi and McCauley, Samuel},
  booktitle={STOC},
  year={2026}
}

@inproceedings{KKMV26,
      title={Preprocessed {3SUM} for Unknown Universes with Subquadratic Space}, 
      author={Yael Kirkpatrick and John Kuszmaul and Surya Mathialagan and Virginia {Vassilevska Williams}},
      year={2026},
      booktitle={ICALP}
}

@inproceedings{GGPS23,
author={Golovnev, Alexander
and Guo, Siyao
and Peters, Spencer
and {Stephens-Davidowitz}, Noah},
title="Revisiting Time-Space Tradeoffs for Function Inversion",
booktitle="CRYPTO",
year="2023",
}

@inproceedings{DG26,
      title={Improved Time-Space Tradeoffs for {3SUM-Indexing}}, 
      author={Itai Dinur and Alexander Golovnev},
      year={2026},
      booktitle={ICALP}
}

@inproceedings{CK19,
  title={The function-inversion problem: Barriers and opportunities},
  author={{Corrigan-Gibbs}, Henry and Kogan, Dmitry},
  booktitle={TCC},
  year={2019},
}

@article{KP19,
  title={The strong {3SUM-INDEXING} conjecture is false},
  author={Kopelowitz, Tsvi and Porat, Ely},
  journal={arXiv:1907.11206},
  year={2019}
}

@article{AESZ23,
  title={Time and space efficient collinearity indexing},
  author={Aronov, Boris and Ezra, Esther and Sharir, Micha and Zigdon, Guy},
  journal={Computational Geometry},
  volume={110},
  pages={101963},
  year={2023},
  publisher={Elsevier}
}

@inproceedings{CEEGS26,
  title={Suffix Random Access via Function Inversion: A Key for Asymmetric Streaming String Algorithms},
  author={Charalampopoulos, Panagiotis and El Ghazi, Taha and Ellert, Jonas and Gawrychowski, Pawe{\l} and Starikovskaya, Tatiana},
  booktitle={ICALP},
  year={2026}
}
\end{document}